\documentstyle[11pt,amsfonts]{article}
\newskip\humongous \humongous=0pt plus 1000pt minus 100pt

\newif\ifdtup

\newcounter{eqnumber}[section]

\makeatletter
\def\@eqnnum{\hbox{\reset@font\rm(\theequation)}}
\let\make@eqnnum=\@eqnnum %
\def\eqnum#1{\dec@eqnnum \global\def\make@eqnnum{\reset@font\rm(#1)}%
\def\@currentlabel{#1}%
}
\def\inc@eqnnum{\addtocounter{equation}{1}}
\def\dec@eqnnum{\addtocounter{equation}{-1}}
\@definecounter{equation}%
\@addtoreset{equation}{section} %
\def\theequation@prefix{{\thesection}.} %
\def\theequation{\theequation@prefix\arabic{equation}}%
\makeatother

\catcode`@=11  
\begin{document}
\begin{titlepage}

\begin{flushright}
cond-mat/0007395 \\
DFUB/00-12\\
September, 2000\\
\end{flushright}

\vskip 2.cm

\begin{center}
{\Large\bf Bose-Einstein condensation in the\\
presence of an impurity}\\

\vspace{1.cm}

{Paola Giacconi$^a$, Fabio Maltoni$^b$, and Roberto Soldati$^a$  }\\
\vspace{.5cm}

$^a$ {\sl Dipartimento di Fisica ``A. Righi'', Universit\`a di
Bologna, and} \\ {\sl Istituto Nazionale di Fisica Nucleare,
Sezione di Bologna, Italia}\\ \vspace{.2cm} $^b$ {\sl Department
of Physics, University of Illinois at Urbana-Champaign,} \\ {\sl
Urbana, IL 61801 USA}

\end{center}

\vspace{2.cm}

\begin{abstract}\noindent
It is shown that Bose-Einstein condensation occurs for an ideal
gas in two spatial dimensions in the presence of one impurity
which is described quantum mechanically in terms of a point-like
vortex and a contact interaction. This model is exactly solvable
and embodies as a special case the analogous problem in three
spatial dimensions.
\end{abstract}

 \vskip 3.0cm

\end{titlepage}

\baselineskip 16pt


\section{Introduction}
\label{IntroSection}

It is well known~\cite{HuangPathria} that Bose-Einstein
condensation, a first order phase transition in momentum space,
can not occur for an ideal gas of free particles in two
dimensions. In this short note we shall show that, by contrast,
the introduction of one point-like impurity on the plane just
allows the Bose-Einstein condensation to take place. Furthermore,
the general pattern of the above phenomenon will be obtained from
a generating Hamiltonian, which encodes a two parameters model to
describe the point-like impurity in two and three spatial
dimensions. The key point to be gathered is how to treat in
quantum mechanics the presence of a point-like, or $\delta$-like,
or null support impurity. Formal manipulations involving some
kinds of regularized $\delta$-like potentials might drive, in
general, to misleading and incorrect conclusions, as the latter
ones are mathematically ill-defined. The correct quantum
mechanical framework to treat point-like
impurities~\cite{Albeverio} is by means of the analysis of the
self-adjoint extensions of the Hamiltonian operator which, in the
present case, turns out to be just a symmetric operator.
Consequently, its domain has to be suitably defined in order to
obtain a self-adjoint Hamiltonian operator which admits a complete
orthonormal set of eigenstates. This construction will be referred
to in the sequel as the inclusion of contact interaction. To be
definite, let us consider as a starting point of our analysis the
following classical one-particle Hamiltonian in two spatial
dimensions: namely,
\begin{eqnarray}
&&H(\phi)={1\over 2m}\left[{\bf p}-{e\over c}{\bf A}({\bf
r})\right]^2\ , \qquad {\bf p}, {\bf r}\in{\bf R}^2\
,\nonumber\\[-6pt]
\label{eq1} \\&&A_j(x_1,x_2)={\phi\over
2\pi}\epsilon_{jk}{x_k\over r^2}\ ,\qquad
r\equiv\sqrt{x_1^2+x_2^2}\,. \nonumber
\end{eqnarray}
Here the Aharonov-Bohm type~\cite{AB} vector potential
corresponds to the presence of a $\delta$-like vortex of flux
$\phi$, which provides a good classical description of one
point-like impurity, as we shall better specify below.
Quantization of the classical Hamiltonian~(\ref{eq1}) leads to a
symmetric operator and, consequently, one has to face the problem
of finding all its self-adjoint extensions. The most general
solution has been recently obtained in Ref.~\cite{AT} and it
consists in a four parameter family. However, to our aim, we can
restrict ourselves to the one parameter sub-family of the $O(2)$
rotational invariant self-adjoint Hamiltonians~\cite{MT,GMS}. The
corresponding spectral decompositions read
\begin{eqnarray}
&& H(\alpha, E_0)=\sum_{l=-\infty}^{+\infty}\int_0^\infty dk\
{\hbar^2 k^2\over 2m}\left|l,k\right>\left<k,l\right| +\vartheta
(-E_0)\left|\psi_B\right>\left<\psi_B\right|\,
,\nonumber\\[-6pt]\label{eq2}\\
&&\alpha\equiv{e\phi\over hc}\in ]-1,0]\ ,\qquad -\infty\le
E_0<+\infty  \nonumber
\end{eqnarray}
$\vartheta$ being the usual Heaviside's step distribution, in
terms of the eigenfunctions
\begin{eqnarray}
\left<r,\theta|l,k\right>={\exp\{il\theta\}\over
\sqrt{2\pi}}\psi_{l}(k,r;\alpha,E_0)\ ,\qquad k\ge 0\ ,
\label{eq3}
\end{eqnarray}
where the improper eigenfunctions belonging to the continuous part
of the spectrum are given by
\begin{eqnarray}
\psi_l(k,r)=\sqrt{k}J_{|l+\alpha|}(kr)\ ,\qquad l\in{\bf Z}-\{0\}\
, \label{eq4}
\end{eqnarray}
\begin{eqnarray}
\psi_0(k,r;E_0)=A(k;\alpha,E_0)J_{|\alpha|}(kr)+B(k;\alpha,E_0)N_{|\alpha|}(kr)\
, \label{eq5}
\end{eqnarray}
in which
\begin{eqnarray}
{B(k;\alpha,E_0)\over A(k;\alpha,E_0)}={\sin(\pi\alpha)\over
\cos(\pi\alpha)+{\rm sgn}(E_0)\left(\hbar^2
k^2/2m|E_0|\right)^\alpha}\ , \label{eq6}
\end{eqnarray}
whereas the normalizable bound state is provided by
\begin{eqnarray}
\left<r|\psi_B\right>=\psi_B(\kappa, r)={\kappa\over
\pi}\sqrt{{\sin(\pi\alpha)\over \alpha}}K_{\alpha}(\kappa r)\
,\qquad \hbar\kappa\equiv\sqrt{2m|E_0|}\ . \label{eq7}
\end{eqnarray}
Some remarks are now in order. First, as previously noticed, the
above spectral decompositions precisely provides the correct
mathematical framework to introduce and properly describe contact
interaction in quantum mechanics. As a matter of fact, it turns
out that the rotational invariant self-adjoint Hamiltonian
operators $ H(\alpha, E_0)$ do represent a one parameter family,
which is labeled by the energy scale $E_0$. In the range $-\infty
<E_0<0$, and only within this range of values, a bound state
$\left|\psi_B\right>$ exists, whose energy is just $E_0$. More
generally, the physical meaning of the characteristic energy scale
$E_0$ is given by the resonance energy, according to the following
pattern: namely,
\begin{eqnarray}
&&E_{\rm res}= \left\{
\begin{array}{ll}
|E_0|({\rm sec} \pi\alpha)^{1/|\alpha|}
 \ , &{\rm if }\, 0<|\alpha|<1/2 \ ,\quad
E_0< 0\ ; \\
|E_0|, &{\rm if }\, 1/2<|\alpha|<1 \ ,\quad E_0< 0\ ;
\end{array}
\right.\\
&&E_{\rm res}=\,\,E_0|{\rm sec} \pi\alpha|^{1/|\alpha|}\ , \qquad\
 {\rm if }\, 1/2<|\alpha|<1\ ,\quad E_0\ge 0\ .
\label{eq8}
\end{eqnarray}
A further observation is that only the non-integer part of the
vortex flux parameter $\alpha$ is actually observable, as its
integer part can always be gauged away by a single valued phase
transformation. To sum up, we can say that a general correct
quantum mechanical description of one point-like impurity is
provided by the two parameters family of self-adjoint Hamiltonians
of Eq.~(\ref{eq2}). The existence of the contact interaction just
corresponds to the presence of a specific locally square
integrable singularity of the wave function at the impurity
position - see Eq.~(\ref{eq5}). In the limit $E_0\to -\infty$
contact interaction is removed, the domain of the Hamiltonian is
that of the regular wave functions on the whole plane and the
impurity is described in terms of a pure Aharonov-Bohm vortex of
non-integer vorticity $\alpha$. If we further take the limit
$\alpha\uparrow 0$, {\it i.e.}, also the Aharonov-Bohm interaction
is turned off, the two dimensional free particle Hamiltonian is
truly recovered (Friedrichs' limit). As we shall discuss in the
sequel, it is curious that just in the Friedrichs' limit the
Bose-Einstein condensation disappears in the two spatial
dimensional case because, in the presence of contact interaction
and no matter how weak it is, a non-vanishing critical temperature
for the Bose-Einstein transition always exists.

\section{One-particle partition function}
\label{sectwo}

In order to discuss Bose-Einstein condensation, it is necessary to
compute the average number of particles at thermal equilibrium. To
this aim, let us first evaluate the diagonal Heat-Kernel and the
one-particle partition function. According to the spectral
decomposition of Eq.~(\ref{eq2}) and after separation of the truly
free particle Hamiltonian $H_0\equiv H(0,-\infty)$ contribution,
it is not difficult to verify that the diagonal Heat-Kernel can be
cast in the following form: namely,
\begin{eqnarray}
G(\alpha,\beta,E_0;r)&\!\equiv\!& G_{\rm
int}(\alpha,\beta,E_0;r)+G_0(\beta)\nonumber\\ &\!=\!& \left<{\bf
r}\left|\left[ \exp\{-\beta H(\alpha,E_0)\} -\exp\{-\beta
H_0\}\right]
\right|{\bf r}\right>+\lambda_T^{-2}\nonumber\\
&\!=\!& I(\alpha;r)+I(-\alpha;r)-2I(0;r)-I_0(\alpha;r)-I_0(-\alpha;r)+I_0(0;r)\nonumber\\
&&+\,\vartheta(-E_0)e^{-\beta E_0}\left|\psi_B(\kappa
r)\right|^2+{\cal I}_0(\alpha,E_0;r) +\lambda_T^{-2}\ ,\label{eq9}
\end{eqnarray}
where the translation invariant free particle diagonal Heat-Kernel
is nothing but the inverse square thermal wavelength
$\lambda_T\equiv (h/\sqrt{2\pi m{\rm k}T})$  and we have set
\begin{eqnarray}
&&I(\alpha;r)= \int_0^{\infty} {dk\over 2\pi}\ k e^{-\beta \hbar^2
k^2/2m} \sum_{l=0}^\infty \; \left [J_{l+\alpha}(kr) \right] ^2\
, \\ &&I_0(\alpha;r)= \int_0^{\infty} {dk\over 2\pi}\ k
e^{-\beta \hbar^2 k^2/2m}\; \left [J_{\alpha}(kr) \right] ^2\ ,\\
&&{\cal I}_0(\alpha,E_0;r)= \int_0^{\infty} {dk\over 2\pi}\
{k\exp\{-\beta\hbar^2 k^2/2m\}\over {1 + {\rm
tan}^2[\pi\mu(k)]+2\, {\rm tan }[\pi\mu(k)] \,{\rm cos} \,
(\alpha\pi)}} \nonumber\\[10pt] &&\hspace*{2.5cm} \times
\left\{\tan^2[\pi\mu(k)] \, J^2_{-\alpha}(kr)+J^2_{\alpha}(kr)+2
\, {\rm tan}[\pi\mu(k)] \,
J_{-\alpha}(kr)\, J_{\alpha}(kr) \right\}\, ,\\[10pt]
&& {\rm tan }[\pi\mu(k)] \equiv {\rm sgn}(E_0) \left[
{2m|E_0|\over \hbar^2 k^2} \right]^{|\alpha|} \,.
\end{eqnarray}
Now, it is
very important to realize~\cite{Virial} that the impurity
interaction part of the diagonal Heat-Kernel $G_{\rm
int}(\alpha,\beta,E_0;r)$ is integrable on the whole plane. This
leads to the following result for the one-particle partition
function
\begin{eqnarray}
Z_{\rm 2D}(\alpha,\beta,E_0)&\!=\!& {{\rm A}\over
\lambda_T^2}+{\alpha(\alpha+1)\over 2} +\vartheta(-E_0)e^{\beta
|E_0|}\nonumber\\ &&\!+ {\alpha\sin(\pi\alpha)\over \pi}
\int_0^{\infty} {dx \over x^{1+\alpha}} {{\rm sgn}(E_0)\ e^{-\beta
|E_0| x} \over 1+ 2{\rm sgn}(E_0)x^{|\alpha|}\cos(\pi\alpha)
+x^{2|\alpha|}}\ ,\label{eq11} \end{eqnarray} where, as usual, we
have denoted by $A$ the area divergence, due to the presence of
the translation invariant part of the free one-particle
Heat-Kernel. The above expression for the one-particle partition
function can be used as a generating form which encodes different
specific notable cases. In particular, the one-particle partition
function in two spatial dimensions and in the presence of pure
contact interaction can be obtained in the limit $\alpha\uparrow
0$ and reads
\begin{eqnarray}
Z_{\rm 2D}(0,\beta,E_B)&=&{{\rm A}\over \lambda_T^2}+ e^{\beta
|E_B|}-\int_0^{\infty} {dE\over E} {e^{-\beta E} \over {{\rm ln}^2
(-E/E_B) + \pi^2}} \\ &=&{{\rm A}\over \lambda_T^2}+\nu(\beta
|E_B|)\ ,\qquad E_B<0\ , \label{eq12}
\end{eqnarray}
where~\cite{Integrals}
\begin{eqnarray}
{\bf \nu}(x)\equiv\int_0^{\infty} {x^t \over {\Gamma (t+1)}} dt\ .
\label{eq13}
\end{eqnarray}
Notice that in two spatial dimensions the bound state is always
present for any $-\infty<E_B<0$. Another distinguished case that
can be read off the basic formula~(\ref{eq11}) is the three
dimensions one-particle partition function in the presence of
contact interaction. As a matter of fact, thanks to dimensional
transmutation~\cite{MT}, the latter case just corresponds to the
value $\alpha=-1/2$, up to a suitable redefinition of the free
part: namely,\\[-5pt]
\begin{eqnarray}
Z_{\rm 3D}(\beta,E_0)={{\rm V}\over \lambda_T^3}
+\vartheta(-E_0)e^{\beta |E_0|} +{1\over 2}{\rm
sgn}(E_0)e^{\beta|E_0|}{\rm erfc}(\sqrt{\beta|E_0|})\ .
\label{eq14}
\end{eqnarray}
\vspace{.0cm}

\section{Results and Discussion}
\label{secthree}

Now we are ready to discuss the Bose-Einstein condensation for an
ideal gas of particles in the presence of one point-like impurity,
as generally described by the one-particle Hamiltonian of
Eq.~(\ref{eq2}). According to the general form~(\ref{eq11}) of the
one-particle partition function, it turns out that the average
particles density  at thermal equilibrium in two spatial
dimensions and in the presence of one point-like impurity is given
by
\begin{eqnarray}
\left<n\right>_{\rm 2D} \equiv {\left<N\right>\over {\rm A}}
&\!=\!& \lambda_T^{-2}g_1(z)+{z\alpha(\alpha+1)\over 2{\rm
A}(1-z)}
+\vartheta(-E_0){z\over {\rm A}(z_0-z)}\nonumber\\
&&+\vartheta(E_0){z\over {\rm A}(1-z)} +{\rm sgn}(E_0){z\over {\rm
A}}\int_0^\infty dE\ {\varrho(E;\alpha,|E_0|)e^{-\beta E}\over
1-z\exp\{-\beta E\}}\ ,\label{eq15}
\end{eqnarray}

\noindent where we have set $z_0\equiv\exp\{\beta E_0\}$ and
$g_1(z)=-\ln(1-z)$, whereas\vspace*{1mm}
\begin{eqnarray}
\varrho(E;\alpha,|E_0|)=
{\alpha\sin(\pi\alpha)E^{|\alpha|-1}\over \pi
|E_0|^{\alpha}\left[E^{2|\alpha|}+|E_0|^{2|\alpha|} +2{\rm
sgn}(E_0)(E|E_0|)^{|\alpha|}\cos(\pi\alpha)\right]}\ .
\label{eq16}
\end{eqnarray}

\noindent It is important to realize that if $E_0<0$ the range of
the fugacity is $0\le z\le z_0<1$, whilst $0\le z\le 1$ if $E_0\ge
0$. Moreover, it is not difficult to prove that, thanks to
analytic continuation, the very last term in Eq.~(\ref{eq15})
admits a finite limit when $z\uparrow 1$ and $E_0\ge 0$. From the
above expression~(\ref{eq15}) for the average particle density in
two spatial dimensions, it appears to be manifest that
Bose-Einstein condensation occurs only in the presence of the
bound state, {\it i.e.}, only for the sub-family of the
self-adjoint extensions of the symmetric Hamiltonian~(\ref{eq1})
in the range $-\infty<E_0<0$. In those cases, the critical
temperature and/or specific area can be obtained as the unique
solutions of the equation
\begin{eqnarray}
\ln\left(1-e^{\beta E_0}\right)=-{h^2\beta\over 2\pi
m}\left<n\right>_{\rm 2D}\ . \label{eq18}
\end{eqnarray}
The three spatial dimensional case can be handled in a quite
similar way, as it essentially corresponds to the specific value
$\alpha=-1/2$ in Eq.~(\ref{eq15}), up to terms irrelevant in the
thermodynamic limit: namely,
\begin{eqnarray}
\left<n\right>_{\rm 3D} \equiv {\left<N\right>\over {\rm
V}}&\!=\!& \lambda_T^{-3}g_{{3\over 2}}(z) +\vartheta(-E_0){z\over
{\rm
V}(z_0-z)} +\vartheta(E_0){z\over {\rm V}(1-z)}\nonumber\\
&&+\, {\rm sgn}(E_0){z\over {\rm V}}\int_0^\infty dE\
{\varrho(E;\alpha=-{1\over 2},|E_0|)e^{-\beta E}\over
1-z\exp\{-\beta E\}}\ . \label{eq19}
\end{eqnarray}
The above equation clearly indicates that Bose-Einstein
condensation always takes place in three spatial dimensions, in
the presence as well as in the absence of the impurity.
Nonetheless, the actual values of the critical temperature and/or
density do depend upon the sign of the parameter characterizing
the self-adjoint extension of the Hamiltonian. In fact, for
$E_0\ge 0$, {\it i.e.} in the absence of the bound state, the
critical values are the usual ones as given by the solution of the
equation $\lambda_T^3\left<n\right>_{\rm 3D}=\zeta(3/2)$. At
variance, when $-\infty<E_0<0$, {\it i.e.} in the presence of the
bound state, the critical values can be read off the equation
\begin{eqnarray}
g_{{3\over 2}}(z_0)=\lambda_T^3\left<n\right>_{\rm 3D}\ .
\label{eq20}
\end{eqnarray}
The case of pure contact interaction in two spatial dimensions can
also be obtained from the basic formula~(\ref{eq15}) taking the
limit $\alpha\uparrow 0$ and treating separately the cases $E_0\ge
0$ and $-\infty< E_0<0$. As a matter of fact, the result is
\begin{eqnarray} \left.\left<n\right>_{\rm
2D}\right|_{\alpha=0}&\!=\!&
\lambda_T^{-2}g_1(z)+\vartheta(E_0){z\over
{\rm A}(1-z)} +\vartheta(-E_0){z\over {\rm A}(z_0-z)}\nonumber\\
&&-\,\vartheta(-E_0){z\over {\rm A}}\int_0^\infty {dE\over E}\
{e^{-\beta E}\over 1-z\exp\{-\beta E\}}{1\over
\ln^2(-E/E_0)+\pi^2}\ ,\label{eq21}
\end{eqnarray}
which shows that condensation does not occur when $E_0\ge 0$,
whereas it appears if $-\infty< E_0<0$, the critical temperature
and/or specific volume being always determined by Eq.~(\ref{eq18})
which does not depend upon $\alpha$. In this latter case, the very
same formula can be also obtained directly from Eq.~(\ref{eq12})
as it does.

In conclusion, we have shown in this note that the presence of
contact interaction makes it possible the occurrence of
Bose-Einstein condensation in two spatial dimensions. This
phenomenon is connected to the presence of a bound state in the
spectrum of the self-adjoint Hamiltonian. It turns out to be
remarkable that the latter circumstance is always there in the
pure contact interaction case, that means without Aharonov-Bohm
vortex interaction. It is in fact worthwhile to notice that, in
the presence of a non-vanishing vorticity $\alpha$, the
half-family without bound state of the self-adjoint extensions
does not allow Bose-Einstein condensation, whilst the remaining
half-family with bound state leads to Bose-Einstein condensation
though the critical temperature is independent from $\alpha$.
Accordingly, only when $-\infty<E_0<0$ a non-vanishing and
vorticity-independent critical temperature is allowed  in the two
spatial dimensional case - see Eq.~(\ref{eq18})- whereas the
critical temperature deviates from its conventional value in the
three spatial dimensional case - see Eq.~(\ref{eq20}).

\end{document}